\documentstyle[prb,aps]{revtex}
\def\be{\begin{equation}}
\def\ee{\end{equation}}
\def\bc{\begin{center}}
\def\ec{\end{center}}
\begin{document}
%\begin{titlepage}
%\bc
%{\Large \bf Critical phenomena at edges and 
%corners}\\[5ex]
\title{
Critical phenomena at edges and corners}
\author{M. Pleimling and W. Selke}
\address{Institut f\"ur Theoretische Physik B, Technische Hochschule,
D--52056 Aachen, Germany}
\maketitle
 
%M. Pleimling and W. Selke\\

%Institut f\"ur Theoretische Physik, Technische Hochschule,
%D--52056 Aachen, Germany \\
%\ec

%\vskip 1.8cm
%\noindent{\Large \bf Abstract}\\[2.5ex]
\begin{abstract}
Using Monte Carlo techniques, the critical behaviour at
edges and corners of the three--dimensional Ising model is
studied. In particular, the critical exponent 
$\beta_2$ of the local magnetization
at edges formed by two intersecting free surfaces is
estimated to be, as a function of the opening angle
$\theta$, $0.96 \pm 0.02$ for $\theta = 135^o$,
$1.28 \pm 0.04$ for $90^o$, and $2.30 \pm 0.10$
for $45^o$. The critical exponent $\beta_3$ of
the corner magnetization of a cube is found to be 
$1.86 \pm 0.06$. The Monte Carlo estimates are compared
to results of mean field theory, renormalization group
calculations and high temperature series expansions.
\end{abstract}

\vspace*{0.5cm}
\noindent
%\pacs{05.50+q, 68.35.Rh, 75.40.Mg}
%\end{titlepage}
 
\section{Introduction}

Critical phenomena may be associated not only with the bulk
of a three--dimensional crystal, but also with its surfaces,
edges and corners. The critical properties of perfect, flat 
surfaces have been studied extensively [1-3].\\
\noindent
However, the singularities of edge and corner quantities have attracted
less attention. Some years ago, Cardy noted and calculated the
dependence of the edge critical exponents on the opening
angle $\theta$ between the surfaces forming the edge (or wedge), using
mean field theory and renormalization group theory of
first order in $\epsilon$ for $O(n)$ models [4]. Most of the
later work dealt with Ising models in two dimensions [5] or
polymers [6]. Rather rarely, the critical behaviour of edges
in three--dimensional Ising models (describing magnetic crystals or 
lattice gas systems such as alloys) has been studied,
applying renormalization group theory or high temperature
series expansions [7-9]. Likewise,
the singular properties at the corners of three--dimensional systems
has been investigated even only in the framework of mean field
theory [5].\\ 
\noindent
Edges and corners may be expected to play a dominant role, for instance, in
nanostructured materials. Recently, many studies have been
performed to reproduce theoretically properties of such small
clusters of atoms, including several Monte Carlo simulations [10,11].
In that context, it looks worthwhile to study systematically
various types of edges and corners in somewhat simplified models, as well, as
will be presented in this article.\\
\noindent
The outline of the article is as follows. In the next section, we shall
consider three--dimensional Ising models with different opening
angles, $\theta$, at the edges. In particular, the critical exponent,
$\beta_2$, of the edge magnetization is estimated, based on
Monte Carlo (MC) data obtained from a cluster--flip algorithm. The values
for $\beta_2(\theta)$ are compared to previous results. In 
section 3, we shall present our Monte Carlo findings on  
Ising cubes, calculating especially 
the corner magnetization and its critical exponent $\beta_3$.
Difficulties in extracting the bulk critical exponent from the
total magnetization of the cubes are pointed out.
A brief summary concludes the article.\\

\section{Edges with different opening angles}

We consider nearest--neighbour Ising models on simple cubic
lattices with ferromagnetic interactions, $J > 0$.
The Hamiltonian is given by
 
\be
H=  - \sum\limits_{(i,j)} J S_i S_j 
\ee
where the sum runs over all bonds $(i,j)$ between spins, $S_i = \pm 1$,
at neighbouring sites $i= (x,y,z)$ and $j$. 
In the thermodynamic limit, the systems display a phase transition
with the magnetization $ \left< S_{xyz} \right>$ vanishing above the critical
point. From numerical analyses, the critical temperature, $T_c$, is
known very accurately, $k_B T_c = 4.511..J$ [12,13].\\
\noindent
To introduce edges in Ising magnets, we apply periodic boundary conditions
along the, say, $z$-axis. The remaining four free surfaces of the crystal
may be oriented in various ways leading to different opening angles $\theta$
at the edges. In particular, for pairs of (100) and (010) surfaces, the
opening angle is $\theta = 90^o$, and the system contains four
equivalent edges; case ({\it a}) in the following (see
Figure 1). Obviously, the
opening angle is invariant against rotating the crystal
about the $z$-axis. However, for
instance, the coordination numbers at the edges may change under a
rotation, and therefore the local thermal edge properties may change. Keeping
$\theta = 90^o$, we rotated the crystal by $45^o$ (yielding (110)
and $(1\overline{1}0)$ free surfaces), in order to check
whether the critical behaviour
at the edges is affected, case ({\it b}). In case ({\it c}), the
surface orientations were chosen to be (100) and (110), with
the intersections forming two pairs of edges with $\theta = 45^o$ and
$\theta = 135^o$. Note that edges with an opening angle of $180^o$
correspond to the free surface, whose critical 
properties have been investigated rather 
extensively before [1,2,14,15].\\
\noindent
In our present Monte Carlo study of Ising magnets with edges, we used
the one--cluster--flip algorithm [16]. Systems with $L \times M \times N$
sites (or spins) were considered, with $L \times M$ being the number
of sites in the planes perpendicular to the $z$--axis, and $N$ being the
number of sites along that axis. Typically, $L = M$ ranged from
20 to 60, with $N$ going up to 640. The concrete values were chosen so that
finite--size effects could be monitored and avoided, when attempting
to elucidate critical properties of indefinitely extended edges, with
the bulk and surface properties being those
of the thermodynamic limit (see below).
Usually, $4 \times 10^4$ clusters were generated per MC run, discarding
the first $10^4$ clusters for equilibration. We averaged over 
an ensemble of at least five runs (using different random numbers)
to obtain the final thermal averages.\\
\noindent
The exchange couplings are assumed to have always the same value, $J > 0$,
i.e. in the bulk as well as at the surfaces and the edges. In the language
of surface critical phenomena [1,2], one then encounters an 
ordinary transition, with the bulk, surfaces and edges ordering at
the same (bulk) critical point, $T_c$. Because one expects
universal critical behaviour at the ordinary transition, that
choice of couplings is merely for reasons of convenience. To cross
over to different universality classes (corresponding to the special point
or the extraordinary and surface transition), the surface couplings
had to be increased significantly [1,2,14,15].\\
\noindent
The crucial quantity of our analysis is the edge magnetization
 
\be
m_2= \frac{1}{N} \left< \left| \sum\limits_{z}  S_{x_0y_0z} \right| \right>
\ee

\noindent
where ($x_0y_0z$) denotes a site at an edge of the lattice (in our
notation, we follow Cardy [4], using the index '2' to indicate that
an edge is formed by the intersection of two planes). In addition, 
we computed the analoguous magnetization for lines parallet to
the $z$--axis, $m_l(x,y)$, as well as the susceptibility, $\chi_2$,
describing the response of the magnetization to a bulk field [4].\\
\noindent
On approach to the bulk critical point, $T_c$, $m_2$ is
expected to vanish as $m_2 \propto t^{\beta_2}$, where $t$ is the
reduced temperature, $t = |T_c - T|/T_c$. To determine $\beta_2(\theta)$, we
define, below and at the critical point, an effective exponent [14] as

\be
\beta_{2,eff}(t)= d \ln (m_2)/d \ln (t)
\ee

\noindent
Because the MC data are recorded at discrete
temperatures, $t_i$, $\beta_{2,eff}$ may be determined numerically by

\be
\beta_{2,eff}(t)= \ln (m(z,t_i)/m(z,t_{i+1}))/\ln (t_i/t_{i+1})
\ee

\noindent
with $t =(t_i + t_{i+1})/2$.\\
\noindent
$\beta_{2,eff}$ is well defined at all temperatures, not merely close
to $T_c$ (for a related recent study of the corresponding critical exponent
near criticality for ferromagnetic thin films, see
Ref. [17]). Obviously, when $t \longrightarrow 0$, $\beta_{2,eff}$ becomes
the asymptotic critical exponent $\beta_2$. In complete analogy, one
may define the effective exponent of the line magnetization, $m_l$. For
sufficiently large systems, $m_l$ approaches deep in
the interior the bulk magnetization, $m_b$, with the corresponding
critical exponent $\beta_b$ = 0.32.. [12,13]. At the surfaces, far from
the edges, $m_l$ approaches the surface magnetization, $m_1$, with the
asymptotic critical exponent being $\beta_1 \approx 0.80$ [14].\\
To estimate reliably $\beta_2$ from the simulational data, finite--size
effects have to be monitored carefully. Strictly speaking, $\beta_2$
is defined in the thermodynamic limit,
$L, M, N \longrightarrow \infty$ (note that we take the absolute
value to define the edge magnetization $m_2$ in (2), because we are
simulating finite lattices). Of course, $m_2$ is affected by spin
fluctuations in the bulk and at the surfaces. Accordingly, the MC
system should be large enough to reproduce the thermodynamic values
of $m_b$ and $m_1$, sufficiently far away from the edge (these
values are known very accurately [12-15]). In general, at a given
temperature and geometry of the crystal, $m_2$ has to be stable
against enlargening the system size, $L \times M$ and $N$.\\
\noindent
In Figure 2, the finite--size effect is illustrated for the geometry of
case ({\it c}), with $\theta = 45^o$ and $135^o$, by plotting
$\beta_{2,eff}$ versus the reduced temperature (error bars,
resulting from ensemble averaging, were determined in the same way
as before [14]). Finite--size dependences, i.e. deviations from
the thermodynamic behaviour, are clearly signalled by a decrease
of $\beta_{2,eff}$ on approach to $T_c$. The crucial quantity is
evidently the length of the edge, $N$ (in the range of temperatures depicted 
in the figure, the size of the planes perpendicular to the $z$--axis 
is sufficiently large, $L= M= 40$, to neglect further finite--size
dependences, as we checked). By comparing our MC data for
$\theta = 45^o, 90^o, 135^o$, and $180^o$ (free surface), an increasing
edge length is needed to circumvent finite--size effects at fixed temperature
and decreasing opening angle.\\
\noindent
This trend may be understood by presuming that the critical behaviour at the
edges is mainly governed by bulk critical fluctuations, similar
to the situation at surfaces [14,15]. Now, by lowering $\theta$, less
pathes between neighbouring spins or sites connect the edges and the
bulk. Accordingly, the edge length has to be enlarged to 'transmit'
the bulk fluctuations fully to the edges at that reduced connectivity.\\
\noindent
It would be of interest to establish the finite--size scaling form
of the edge magnetization $m_2(t, L=M, N)$ or, more generally, of
the line magnetization $m_l(\vec{d}, t, L=M, N)$, where $\vec{d}$ denotes
the distance of the line to the edge. However, such analyses are 
beyond the scope of this article.--It seems worth mentioning that our
MC data suggest that the profile $m_l(\vec{d})$ approaches the surface
magnetization, $m_1$, or the bulk magnetization, $m_b$, along, e.g., the 
shortest pathes, in an exponential form, i.e.   
 
\be
m_{b(1)}- m_l(\vec{d}) \propto exp (-a_{b(1)} \left| \vec{d} \right|).
\ee

\noindent
From a preliminary analysis of our MC data, it seems
conceivable that $a_b$ and
$a_1$ become identical on approach to $T_c$, being related to the bulk
correlation length. A similar behaviour holds for the magnetization
profile as a function of the distance from the surface [14,15,18,19].\\
\noindent
Figure 3 summarizes our simulational results for the effective exponent
$\beta_{2,eff}(\theta,t)$ in the cases ({\it a}) and ({\it c}), i.e.
with one pair of (100) surfaces. Only those MC data are shown which
are not significantly affected by finite--size effects. We also included
results of our previous MC study for the surface
magnetization [14], corresponding to $\theta =180^o$.\\
\noindent
The slope of $\beta_{2,eff}$ changes rather mildly at sufficiently
small reduced temperatures $t$. Thereby meaningful estimates for
the asymptotic exponent $\beta_2(\theta)$ seem to be feasible, with
uncertainties rising when lowering the opening angle, because of a lack
of MC data close to criticality due to the stronger
finite--size effect. Our estimates are $\beta_2(\theta=180^o)= 0.80 \pm
0.01$ [14], $\beta_2(135^o) =0.96 \pm 0.02$, $\beta_2(90^o)= 1.28 \pm 0.04$,
and $\beta_2(45^o)= 2.30 \pm 0.10$, with error
bars referring to 'reasonable' extrapolations of the effective exponents
to the critical point.\\
\noindent
These MC estimates are to be compared with results obtained from
mean field theory (MF) [4], renormalization group calculations
to first order in $\epsilon$ (RNG) [4], and high temperature series
expansions (HTS) [7], see Figure 3 and Table 1. The values 
given by mean field theory are presumably systematically too high, according
to the predictions of the renormalization group. These predictions, in
turn, seem to be systematically too large, as suggested both by high
temperature series expansions and the MC simulations. The last two
methods yield results which are in close agreement with each other. A similar
tendency in the values of $\beta_2(\theta)$ obtained from the different
analytical and numerical approaches (MF, RNG, HTS, and MC) holds
for the $n$--vector model with
$n = 0$, describing polymers [4,7,20].\\
\noindent
At the fixed opening angle $\theta = 90^o$, we studied the
effect of a rotation of the crystal
on the critical edge properties, case ({\it b}). In particular, the
crystal was rotated by $45^o$ about the $z$--axis. The
wedge is then formed by (110) and $(1\overline{1}0)$
surfaces. By the rotation, the coordination numbers at
the surfaces and edges are
reduced by one. Consequently, the surface and edge magnetizations are
lowered. Following the above considerations, the finite--size effect
is expected to be enhanced compared to case ({\it a}), i.e. after
rotation, larger MC systems are needed to approximate closely the
thermodynamic behaviour. The MC data confirm this expectation.\\
\noindent
Results on the effective exponent $\beta_{2,eff}$ are displayed in
Figure 4, comparing the cases ({\it a}) and ({\it b}). Again, we 
depict only MC data which are not significantly affected by
finite--size effects. It seems well
conceivable to extrapolate both cases to
the same asymptotic critical exponent, $\beta_2 = 1.28 \pm 0.04$, consistent
with the invariance of the value of the boundary critical exponent against
rotation of the crystal. After rotation, we
did not approach the critical point as closely as before, because otherwise
extremely large systems had to be dealt with. In any event, as seen
in Figure 4, corrections to scaling are much weaker in case ({\it b}), for
small reduced temperatures, resulting in the much lower slope
of $\beta_{2,eff}$, compared to Ising models with (100) and (010)
surfaces.\\
\noindent
Note that the invariance of boundary critical exponents against rotation
of the crystal has been discussed and partly even
proven for surfaces and edges of 
two--dimensional Ising models [5,19,21].\\
\noindent
Our MC data for the exponent of the edge susceptibility, $\chi_2$, are
less accurate than those for the
edge magnetization, thereby allowing to estimate the corresponding
critical exponent $\gamma_2$ only with rather large uncertainties.
At any rate, $\gamma_2$ is found to depend strongly on the opening
angle $\theta$, and the results of the RNG calculations 
and HTS method [4,7] seem to provide fairly good estimates
for the true values. \\

\section{Cubes and corners}

We consider Ising cubes of $L^3$ spins on simple cubic
lattices with free surfaces. In the thermodynamic
limit, $L \longrightarrow \infty$, the corner magnetization is given by
 
\be
m_3= \frac{1}{8} \left<  \sum\limits_{x_0y_0z_0}  S_{x_0y_0z_0}  \right>
\ee

\noindent
summing and averaging over the eight corner spins at site $(x_0y_0z_0)$. In
complete analogy to (3), one may define an effective exponent
$\beta_{3,eff}$, approaching the asymptotic corner critical exponent
$\beta_3$, as $t \longrightarrow 0$.\\
\noindent
In a finite system, $m_3$ is, strictly speaking, zero, because the
Hamiltonian is invariant against reversing the spins. On the other
hand, for instance, the absolute corner magnetization
 
\be
m_{3,a}= \frac{1}{8} \left< \left| \sum\limits_{x_0y_0z_0}  S_{x_0y_0z_0} \right| \right>
\ee

\noindent
seems to show no critical
behaviour. Indeed, $m_{3,a}$ is quite distinct
from $m_3$, involving a fixed number of spins independent of the
system size, in contrast to the bulk, surface or edge magnetizations
calculated with or without absolute values.\\
\noindent 
To compute $m_3$, we use the standard one--spin--flip MC algorithm (Metropolis
algorithm). Taking the fully ordered ground state with, say, $S_{xyz} = 1$
at all sites as starting configuration of the simulation, one 
encounters, after initial relaxation, at $T < T_c$, a
metastable state, in which the total magnetization is positive. The
system remains in that metastable state (being separated by an
energy barrier from the 'mirror' metastable state with negative
total magnetization), possibly, for a long time, the time
depending on the size $L$ and the temperature, see, e.g., Ref. [22]. The
corresponding local magnetizations, especially the
corner magnetization $m_3$, do not vanish then, and their values, taking usual
finite--size dependences into account, are supposedly very close to those
in the thermodynamic limit.\\
\noindent
In fact, by monitoring the total magnetization during the 
MC runs, we identified the 
temperatures at which we can reliably compute $m_3$. In Figure 5, resulting
findings for the effective exponent, $\beta_{3,eff}$, are shown. Most of the
data are for rather small cubes (mimicking magnetic nanoparticles), with 
$L = 20$. 3000 Monte Carlo steps per site (MCS)
were used to 'equilibrate' (to the metastable state) the systems;
additional 7000 MCS were performed to obtain thermal averages. To
improve the statistics, we sampled over an ensemble of a few 
$10^2$ independent MC runs at
each temperature. No finite--size effects were observed, when we increased
$L$ to 40 at the smallest reduced temperatures, see Figure 5. To generate data
of the high accuracy required to determine reliably
the effective exponent without finite--size corrections 
closer to $T_c$, one had to simulate eventually
much larger system sizes. In that way, the computing
time had to be increased appreciably then. We refrained from doing so, because
the present data already allow for a decent estimate of the asymptotic
critical exponent, $\beta_3 = 1.86 \pm 0.06$. That value is significantly
lower than the one predicted by mean field theory, $\beta_3 =2$ [5].\\
\noindent
In finite--size scaling theories [23,24], one usually considers the total
magnetization of the system, summing and averaging over all spins.
This approach may work in the simplest form by using periodic
boundary conditions, implying translational invariance. Of course, in
the case of free surfaces much care is needed, because the
contributions of bulk, surface, edge, and corner spins to the total
magnetization are of different type. Actually, the local magnetization
differs, in general, from site to site.\\
\noindent
In Figure 6, the effective exponent, $\beta_{T,eff}$ of the total
magnetization $m_T$, taking, in analogy to (2), the absolute values,
is shown again for a relatively small cube with $20^3$ spins (including, for
comparison, data for $L = 40$). $\beta_{T,eff}$ is appreciably
higher than the bulk critical
exponent $\beta_b = 0.32..$ already quite far away from
criticality, $t < 0.2$. The 'overshooting
phenomenon' is caused by the spins near the surfaces, edges and
corners, reducing the total magnetization and 
yielding large effective and critical exponents, as
discussed in this article. Certainly, on approach to $T_c$ the
effective exponent goes to zero, because
the absolute total magnetization $m_T$ does not
vanish at the critical point in a finite system.\\
\noindent
The overshooting persists at larger system sizes (becoming,
however, weaker at fixed $t$; see Figure 6 for $L$ =40), and it may be
difficult to extract the correct bulk critical
exponent reliably. In an alternate
approach, one may replace the reduced temperature $t$
by $t_m= |T - T_m|/ T_m$, where $T_m$ denotes the turning point of the
total magnetization for a given system size, $L$. From
$\beta_{eff}$ = $d$ln $(m_T)$/$d$ln $(t_m)$, one might estimate 
$\beta_b$ (for $L \longrightarrow \infty$ one has $t_m \longrightarrow t$,
and thence $\beta_{eff} \longrightarrow \beta_b$ as $t \longrightarrow 0$) in
an easy way, see Figure 6. For larger system sizes, $\beta_{T,eff}$ and
$\beta_{eff}$ tend to approach each other, as one observes, e.g., for systems
with $L$ = 40. In any event, detailed considerations would
be desirable to substantiate and quantify the finite--size effects for Ising
cubes (for related problems, demonstrating the sensitivity of 
finite--size scaling to different boundary conditions, see, e.g.,
Ref. [23,24]).

\section{Summary}

Using Monte Carlo techniques, we studied critical properties of edges
and corners in three--dimensional Ising models. In particular, by
computing effective exponents for the edge and corner magnetization,
the corresponding asymptotic critical exponents could be estimated
quite reliably.\\  
\noindent
The critical exponents at the edges are found to depend sensitively
on the opening angle formed by the surfaces intersecting at the edge.
The numerical values for these exponents, based on the
simulational data, refined predictions of renormalization group
calculations of first order in $\epsilon$, and confirmed results
of high temperature series expansions. The exponents seem to be
invariant against rotating the crystal, but keeping the opening angle,
as we demonstrated in a specific case. However, the corrections
to scaling may be largely affected by such a rotation.\\  
\noindent
The critical exponent at the corner had been calculated before
only by using mean field theory. The estimate based on the
Monte Carlo data shows that the true value is significantly lower
than the one predicted by mean field theory. It would be quite
interesting to apply other numerical and analytical techniques
to determine that exponent.\\
By simulating corner properties, we considered Ising models
on a cube with free surfaces (mimicking magnetic
nanoparticles). Such systems show very strong
finite size effects, when calculating quantities averaging
over the entire lattice. Care is needed in extracting
bulk critical properties, because the singularities
at the surfaces, edges, and corners are distinct from the
bulk singularities, and play an important role even for
fairly large system sizes.\\[1cm]  

\noindent{\large \bf Acknowledgements}\\[2ex]
 
We should like to thank H. W. Diehl for suggesting to us to study
the edge problem, and F. Igl\'{o}i for a 
useful discussion. It is a pleasure, to thank here
J. Zittartz for gratifying interactions 
with one of us (W.S.) during many years in organizing the
Landau--Seminars 'Cooperative Phenomena in Many--Particle Systems
in Physics'.\\

\vskip 1.3cm
\bc
{\Large \bf References}\\[2ex]
\ec
\begin{enumerate}
%1
\item Binder, K.: In: Phase Transitions and Critical Phenomena. Vol.8,
Domb, C. and Lebowitz, J.L. (eds). London: Academic Press 1983
%2
\item Diehl, H.W.: In: Phase Transitions and Critical Phenomena. Vol.10,
Domb, C. and Lebowitz, J.L. (eds). London: Academic Press 1986; In:
Proceedings of Third International Conference 'Renormalization Group-96'.
Singapore: World Scientific (to be published)
%3
\item Dosch, H.: Critical Phenomena at Surfaces and Interfaces. Berlin,
Heidelberg, New York: Springer 1992
%4
\item Cardy, J.: J.\ Phys.\ A: Math.\ Gen.\ {\bf 16}, 3617 (1983)
%5
\item Igl\'{o}i, F., Peschel, I., Turban, L.: Adv.\ Phys.\ {\bf 42},
683 (1993)
%6
\item De Bell, K., Lookman, T.: Rev.\ Mod.\ Phys.\ {\bf 65}, 87 (1993)
%7
\item Guttmann, A.J., Torrie, G.M.: J.\ Phys.\ A: Math.\ Gen.\ {\bf 17},
3539 (1984)
%8
\item Wang, Z.-G., Nemirovsky, A.M., Freed, K.F., Myers, K.R.:
J.\ Phys.\ A: Math.\ Gen.\ {\bf 23}, 2575 (1990)
%9
\item Larsson, T.A.: J.\ Phys.\ A: Math.\ Gen.\ {\bf 19}, 1691 (1986)
%10
\item Merikoski, J., Timonen, J., Manninen, M., Jena, P.:
Phys. Rev. Lett. {\bf 66}, 938 (1991)
%11
\item Dimitrov, D.A., Wysin, G.M.: Phys. Rev. B {\bf 54}, 9237 (1996)
%12
\item Ferrenberg, A.M., Landau, D.P.: Phys. Rev. B {\bf 44}, 5081 (1991)
%13
\item Talapov, A. L., Bl\"ote, H. W. J.: J. Phys. A: Math. Gen. {\bf 29}, 5727
(1996)
%14
\item Pleimling, M., Selke, W.: Eur. Phys. J. B {\bf 1}, xxx (1998)
%15
\item Landau, D.P., Binder, K.: Phys. Rev. B {\bf 41}, 4633 (1990)
%16
\item Wang, J. S., Swendsen, R. H.: Physica A {\bf 167}, 565 (1990);
 Wolff, U.: Phys. Rev. Lett. {\bf 60}, 1461 (1988)
%17
\item Schilbe, P., Rieder, K.H.: Europhys. Lett. {\bf 41}, 219 (1998)
%18
\item Wingert, D., Stauffer, D.: Physica A {\bf 219}, 135 (1995)
%19
\item Selke, W., Szalma, F., Lajko, P., Igl\'{o}i, F.: J. Stat. Phys. {\bf 89},
1079 (1997);
Igl\'{o}i, F., Lajko, P., Selke, W., Szalma, F.: J. Phys. A: Math. Gen.
 (in print)
%20
\item Gaunt, D.S., Colby, S.A.: J. Stat. Phys. {\bf 58}, 539 (1990)
%21
\item Barber, M.N., Peschel, I., Pearce, P.A.: J. Stat. Phys. {\bf 37},
497 (1984); Peschel, I.: Phys. Lett. {\bf 110A}, 313 (1985)
%22
\item Miyashita, S., Takano, H.: Prog. Theoret. Phys. {\bf 73},
1122 (1985)
%23
\item Barber, M.N: Phase Transitions and Critical Phenomena. Vol.8,
Domb, C. and Lebowitz, J.L. (eds). London: Academic Press 1983;
Fisher, M. E.: In: 'Critical Phenomena', Proc. 51st Enrico Fermi Summer
School, M. S. Green (ed.). New York: Academic Press 1970
%24
\item Binder, K.: In: 'Computational Methods in Field Theory', H. Gausterer,
 C.B. Lang (eds.). Berlin: Springer 1992; Dohm, V.: Physica
Scripta T {\bf 49}, 46 (1993); Dasgupta, S., Stauffer, D., Dohm, V.: Physica
A {\bf 213}, 368 (1995)
\end{enumerate}

\begin{table}
\caption{Predictions for edge critical exponent $\beta_2$, using various
methods discussed in the text.}
\begin{tabular}{|c||c|c|c|c|}
\hline
 & $45^o$ & $90^o$ & $135^o$ & $180^o$ \\
\hline\hline
MF [4] & 2.50 & 1.50 & 1.17 & 1.00 \\
\hline
RNG [4] & 2.48 & 1.39 & 1.02 & 0.84 \\
\hline
HTS [7] & 2.30 & 1.31 & 0.98 & 0.81 \\
\hline
MC & $2.30 \pm 0.10$ & $1.28 \pm 0.04$ & $0.96 \pm 0.02$ & $0.80 \pm 0.01$ [14]\\
\hline
\end{tabular}
\end{table}

\bc
{\Large \bf Figure Captions}\\[2ex]
\ec
\begin{itemize}
\item[Fig. 1:] Geometry of an Ising model with (100) and (010) surfaces
(shadowed), i.e. edges with opening angle $\theta =90^o$.
 
\item[Fig. 2:] Effective exponent $\beta_{2,eff}$ of the edge magnetization
versus reduced temperature, $t$, at edges with opening angles $\theta$ =
(a) $45^o$ and (b) $135^o$, simulating different sizes $L \times M \times N$,
see inset.
 
\item[Fig. 3:] $\beta_{2,eff}$ versus $t$ at edges with opening 
angles $\theta$ = $180^o$ (triangles)[14], $135^o$ (diamonds), $90^o$
 (squares), and $45^o$ (circles). MC system sizes were chosen to
circumvent finite--size effects, see text. The predictions of RNG 
calculations (dashed line)[4] and HTS expansions (dot--dashed line) [7] for
$\beta_2$ are shown as well.

\item[Fig. 4:] $\beta_{2,eff}$ versus $t$ at edges
with $\theta = 90^o$, with pairs of (100) and (010) surfaces
(open circles) and pairs of (110) and $(1\overline{1}0)$ surfaces
 (full circles), for MC system sizes up to $60 \times 60 \times 640$
to avoid finite--size dependences.

\item[Fig. 5:] Effective exponent $\beta_{3,eff}$ of the corner
magnetization as a function of $t$ for Ising cubes with $20^3$ spins
(open circles) and, for comparison at small $t$, with $40^3$
spins (full triangles).
The broken line is a guide to the eye. The solid line refers to
the value of $\beta_3$ as calculated from mean field theory [5]. 

\item[Fig. 6:] Effective exponent of the total
magnetization $\beta_{T,eff}$ ($\beta_{eff}$; full symbols) versus
reduced temperature $t$ ($t_m$) of an Ising cube with $20^3$
(squares) and $40^3$ (circles) spins, using
different reference temperatures for the (finite--size) critical point
$T_c$ ($T_m$), see text. 

\end{itemize}
\end{document}